\documentclass{article}
\usepackage{amsmath,graphicx,mlspconf}
\usepackage{amssymb,amsfonts}
\usepackage{epsfig}
\usepackage{adjustbox}
\usepackage{cite}
\usepackage{todonotes}
\usepackage{hyperref}
\usepackage{booktabs}
\usepackage{graphicx}
\usepackage{float}
\newcommand{\comment}[1]{}
\newcommand{\B}{\color{blue}}
\newcommand{\R}{\color{red}}
\newcommand{\G}{\color{green}}
\comment{
\let\oldbibliography\thebibliography
\renewcommand{\thebibliography}[1]{%
  \oldbibliography{#1}%
  \setlength{\itemsep}{0pt}%
}}

\usepackage[skip=5pt]{caption}

\hypersetup{
    colorlinks=true,
    linkcolor=olive,
    citecolor=magenta,      
    urlcolor=cyan,
    pdftitle={NAaLoss}
    }

\copyrightnotice{979-8-3503-2411-2/23/\$31.00 {\copyright}2023 IEEE}

\toappear{2023 IEEE International Workshop on Machine Learning for Signal Processing, Sept.\ 17--20, 2023, Rome, Italy}

\title{NA\MakeLowercase{a}Loss: Rethinking the Objective of Speech Enhancement}
\name{Anonymous}
\address{Anonymous}

\name{%
    Kuan-Hsun Ho$^{\dagger}$%
    \qquad En-Lun Yu$^{\dagger}$
    \qquad Jeih-weih Hung$^{\star}$%
    \qquad Berlin Chen$^{\dagger}$%
}
\address{%
    $^{\dagger}$ \small{Department of Computer Science and Information Engineering, National Taiwan Normal University, Taipei, Taiwan} \\%
    $^{\star}$ \small{Department of Electrical Engineering, National Chi Nan University, Nantou, Taiwan}
}

\begin{document}

\maketitle

\begin{abstract}
Reducing noise interference is crucial for automatic speech recognition (ASR) in a real-world scenario. However, most single-channel speech enhancement (SE) generates "processing artifacts" that negatively affect ASR performance. Hence, in this study, we suggest a Noise- and Artifacts-aware loss function, NAaLoss, to ameliorate the influence of artifacts from a novel perspective. NAaLoss considers the loss of estimation, de-artifact, and noise ignorance, enabling the learned SE to individually model speech, artifacts, and noise. We examine two SE models (simple/advanced) learned with NAaLoss under various input scenarios (clean/noisy) using two configurations of the ASR system (with/without noise robustness). Experiments reveal that NAaLoss significantly improves the ASR performance of most setups while preserving the quality of SE toward perception and intelligibility. Furthermore, we visualize artifacts through waveforms and spectrograms, and explain their impact on ASR. \comment{To the best of our knowledge, this is the first work that reflects upon past objective functions in SE and proves the existence of artifacts. }
\end{abstract}
\section{Introduction}
\label{sec:intro}

The goal of speech enhancement (SE) is to enhance the quality and intelligibility of a speech signal contaminated by various kinds of noise. Recent advances in machine learning and deep neural networks have shown promise in improving SE performance by learning to model the complex statistical relationships between clean speech and noise. In particular, many studies have shown that multi-channel SE behaves well, even if cascaded with automatic speech recognition (ASR) systems \cite{mc1, mc2}. However, multiple-array microphones are still uncommon, and single-channel SE is essential in real-world scenarios such as mobile devices and hearing aids.

It is noteworthy that while single-channel SE help reduces the impact of noise on speech signals, it may also introduce unnatural artifacts and distortions \cite{badse1, badse2, badse3, artif}. These issues could result in mistakes during the feature extraction stage of the ASR system, which relies on accurate representations of the speech signal for classification. For example, suppose an SE algorithm introduces artifacts that alter the timing or duration of speech sounds. It could lead to misaligned words or phonemes, which lowers ASR performance. The discrepancy in training objectives between SE and ASR could be a reason for the presence of artifacts. Although joint training \cite{badse2, jtr} or data-augmentation techniques \cite{badse3, daug} have been proposed to address this issue, revising the ASR system is not always practical. For SE-based robust ASR, it is crucial to eliminate the detrimental artifacts that hinder recognition. 

Since artifacts differ depending on the SE models used, finding an explicit expression that formulates artifacts is arduous. One endeavor along this direction is orthogonal projection-based error decomposition \cite{artif, ortho}, which analyzes signals by projecting them into the space occupied by speech and noise. However, these formulations are somewhat inaccurate because noise and source signals may not be orthogonal, as in the babble noise scenario.

Is there a facile way to train an SE model that not only maintains the already established enhancement quality but also improves recognition accuracy? We reckon that the solution lies in the employed objective function. Typically, the loss function in SE minimizes the distance between the estimated and target clean speech \cite{typloss1, typloss2}. However, the learning scheme that distinguishes between speech and noise signals has yet to be comprehensively considered, making SE models unable to discern between the speech and noise components of a noisy speech. In particular, the mapping-based SE methods tend to produce false alarms or fake speech\footnote[1]{as shown in supplementary file \url{https://reurl.cc/01aOYY}, under directory \textit{"wavs/false\_alarm"}.}, as their initial design is not meant to separate speech and noise. Consequently, we only focus on masking-based SE and attempts to upgrade them.

We rethink the goal of SE and outline four expectations. For a masking-based SE front-end with a noisy input:
\begin{itemize}
\itemsep 0em 
\item[1.]the output should not include artifacts;
\item[2.]the noise component should be masked out while retaining clean speech;
\item[3.]speech quality and intelligibility should be optimized;
\item[4.]the SE should benefit the subsequent ASR of any form.
\end{itemize}
Accordingly, we present a Noise- and Artifact-aware loss function, NAaLoss, to learn the SE framework, aiming to bridge the gap between reality and the above expectations. An extensive set of experiments exhibit that the presented NAaLoss benefits the SE method by providing significant improvement toward ASR, demonstrating its superiority.

\comment{
The remainder of the paper is composed as follows. Section~\ref{sec:formulation} explains the problem formulation along with the proposed solution. Then in Section~\ref{sec:exp}, we detail the experiments and explain the results. Discussion and case studies are presented in Section~\ref{sec:dis_cs}. Finally, the conclusion is drawn in Section~\ref{sec:conclu}.
}

\section{Formulation}
\label{sec:formulation}

\subsection{Problem}
To propose a comprehensive solution, we must first identify the problem. Despite their diversity and unpredictability, artifacts can be characterized as signals that 1) degrade the Word Error Rate (WER), 2) are ignored in perception/intelligibility metrics, 3) are produced by the SE module, and 4) are distortions to natural signals.\cite{artif} We can then narrow down how the artifacts can be formulated using these characteristics.

A noisy speech $z \in \mathbb{R}^{T}$ can be modeled through a single microphone as $z = x + y$, where $x \in \mathbb{R}^{T}$ is the clean speech, and $y \in \mathbb{R}^{T}$ is the noise. Let $f$ denote a SE model, and $\theta$ denotes artifacts. We hold three hypotheses:
\begin{itemize}
\itemsep 0em 
\item[1.] $f(x) = \theta_c + x$; an SE model $f$ consumes clean speech $x$ and outputs a combination of clean-conditioned artifacts $\theta_c$ and clean speech $x$.
\item[2.] $f(z) = \theta_m + \tilde{y} + x$; an SE model $f$ consumes noisy speech $z$ and outputs a mixture of multi-conditioned artifacts $\theta_m$, residual noise $\tilde{y}$, and clean speech $x$.
\item[3.] $f(y) = \tilde{y}$; residual noise $\tilde{y}$ is the outcome of an SE model $f$ fed with pure noise $y$.
\end{itemize}

Inspecting previous works, most do not explicitly define artifacts introduced by non-linear transformations in SE models. From these hypotheses above, we can formulate artifacts using two options:
\begin{itemize}
\item[$\alpha$.] $\theta = \frac{1}{2}(f(z)+f(x)-f(y)-2x)$, referred to as condition-invariant artifacts by assuming $\theta_c = \theta_m = \theta$.
\item[$\beta$.] $\theta_c = f(x)-x$ and $\theta_m = f(z)-f(y)-x$, which models clean- and multiple-conditioned artifacts individually.
\end{itemize}

While option $\alpha$ is straightforward and the artifact term can be determined by merely solving the equations in the three hypotheses, it may not always be feasible. Contrarily, option $\beta$ asserts that artifacts are case-sensitive, and that is more realistic. Moreover, the characteristics of the received signal always change in a real environment, making it necessary to model artifacts created by an SE module from a multi-aspect perspective.

\subsection{Proposed solution}

Generally, the loss function utilized in an SE module is the distance between the enhanced and target speech representations\cite{typloss2}. However, as previously indicated, artifacts generated by SE regarding different inputs have yet to be considered. As a consequence, we propose the Noise- and Artifacts-aware Loss function (NAaLoss), which contains three components as follows:
\begin{itemize}
\itemsep 0em 
\item[1.] Loss of estimation, $L_\text{estim} = \text{dist}(f(z), x)$, which calculates the distance between the enhanced speech $f(z)$ and target clean speech $x$. This loss is employed in the learning of most SE models.
\item[2.] Loss of de-artifact, $L_\text{deatf}$. In particular, $L_\text{deatf} = \text{dist}(\theta, \mathbf{0})$ if we use option $\alpha$, which treats artifacts to be condition-invariant. On the contrary, $L_\text{deatf} = \sum_i \text{dist}(\theta_i, \mathbf{0}), i \in \{c,m\}$ to consider artifacts coming from different conditions as in option $\beta$. This loss considers artifacts to learn an SE model.
\item[3.] Loss of noise ignorance, $L_\text{ignor} = \text{dist}(f(y), \mathbf{0})$, which measures the magnitude of residual noise $\tilde{y}$. This loss reflects SE's capacity for noise reduction.
\end{itemize}

Here, the symbol $\mathbf{0}$ indicates a tensor filled with zeros, and the function $\text{dist}(\cdot)$ can be any type of distance metric, such as L1 or L2, performed on any feature domain of signals.
Afterward, we build the NAaLoss as a weighted sum of these three components:
$$
L_{\text{NAa}} = (1-\alpha-\beta) L_\text{estim} + \alpha L_\text{deatf} + \beta L_\text{ignor},
$$
where $\alpha=0.1$ and $\beta=0.1$ are designated empirically.
\comment{$\alpha=||\theta||^2_2 / (||z||^2_2+||\theta||^2_2+||y||^2_2)$ and $\beta=||y||^2_2 / (||z||^2_2+||\theta||^2_2+||y||^2_2)$}

We evaluate our proposed loss function on two SE models, one simple and one advanced. Each model is either pre-trained ($pre$) or trained-from-scratch ($scr$). The pre-trained type, in particular, implies that the initial parameter set has been pre-optimized for gaining the best speech quality and intelligibility. \comment{A schematic diagram is illustrated in Fig~\ref{fig:pre}.}
As for the two SE models, the simple one is set to be the example recipe in Speechbrain \cite{sb}, whereas MANNER \cite{manner} is chosen as the advanced one. Both models are masking-based. However, the former operates in the time-frequency domain, and the latter directly handles time signals. With this setting, we can also test the generalizability of the proposed loss function.

\comment{
\begin{figure}[tb]
\centerline{\includegraphics[width=6cm]{NAaLoss/pretrain.png}}
\caption{A schematic diagram illustrating the pretrained process.}
\label{fig:pre}
\end{figure}
}

Finally, we generate 7 combinations of models. The original simple/advanced SE models without further learning are denoted as $f^{S/A}_{org}$, and the simple/advanced models further learned through NAaLoss option $\alpha$/$\beta$ from pre-trained parameters/trained-from-scratch are denoted in the form of $f^{S/A}_{pre/scr, \alpha/\beta}$. For example,  $f^A_{pre, \alpha}$ describes the advanced model with pre-trained parameters that is further learned with NAaLoss using option $\alpha$.

\section{Experiments}
\label{sec:exp}

\subsection{Experimental Settings}

To validate the effectiveness of our proposed solution, we conduct a series of experiments on the VoiceBank-DEMAND \cite{vcbk-dm} benchmark dataset, a widely-adopted and open-source benchmark corpus for SE. In the training set, 11,572 utterances (from 28 speakers) are pre-synthesized with 10 types of noise from the DEMAND database \cite{dm} at four different signal-to-noise ratio (SNR) values: 0, 5, 10, and 15 dB, while the test set contains 824 utterances (from 2 speakers) contaminated by five types of noise at SNR values of 2.5, 7.5, 12.5, and 17.5 dB. In addition to training, we set aside around 200 utterances from the training set for validation. All speech data have a sample rate of 16 kHz.

The configuration of simple SE remains unchanged, as provided in the repository\footnote[2]{\url{https://github.com/speechbrain/speechbrain/tree/develop/recipes/Voicebank/enhance/spectral_mask}}. On the other hand, we use MANNER-small for advanced SE along with the configuration default in the repository\footnote[3]{\url{https://github.com/winddori2002/MANNER}}. We optimize SE models using the Adam optimizer \cite{adam} with past momentum loaded. \comment{The momentum helps the optimizer to continue moving in the same direction as the previous updates, which helps to avoid getting stuck in local minima.}

To see whether NAaLoss alleviates artifacts under any form of subsequent ASR, as outlined in Section~\ref{sec:intro}, we prepare two ASR systems, CCT-AM and MCT-AM, to recognize the enhanced speech. The CCT-AM denotes the acoustic model (AM) trained in the clean condition mode using all the clean utterances. \comment{in VoiceBank \cite{vcbk}.} MCT-AM, on the other hand, indicates the AM trained with utterances corrupted by multi-condition noises. \comment{The noisy utterances are synthesized by adding the clean recordings with 13 types of noise sources from DEMAND with SNR values ranging from 0 to 15 dB. All utterances in the test set differ from the training set in speaker, utterance content, and noise type.} Both AMs are Kaldi-based hybrid DNN-HMM acoustic models \cite{kaldi} trained with lattice-free MMI objective function, and their DNN components utilize time-delay neural networks. 

\subsection{Baselines}

\begin{table}[tb]
\caption{The baselines. The header indicates the input to ASR, while the first and second rows show the WERs (lower is better) using the ASR first column indicated. The PESQ and STOI scores (greater is better) are in the third and fourth rows. CCT/MCT-AM denotes the acoustic model trained in the clean/multi condition mode.}
\label{tab:baseline}
\resizebox{\columnwidth}{!}{%
\begin{tabular}{@{}ccccccc@{}}
\toprule
         & $x$  & $f^S_{org}(x)$ & $f^S_{org}(z)$ & $f^A_{org}(x)$ & $f^A_{org}(z)$ & $z$   \\ \midrule
CCT-AM   & 5.04 & 5.97           & 10.21          & 5.28           & 7.37           & 23.76 \\
MCT-AM   & 4.86 & 5.10           & 7.21           & 4.91           & 6.62           & 8.32  \\ \midrule
PESQ     & -    & 4.47           & 2.69           & 4.22           & 3.11           & 1.97  \\
STOI(\%) & -    & 99.6           & 93.8           & 98.8           & 94.9           & 92.0  \\ \bottomrule
\end{tabular}%
}
\end{table}

Tab.~\ref{tab:baseline} shows the results of various SE and ASR baselines, including clean speech $x$, noisy speech $z$, and enhanced speech with respect to simple or advanced SE models. It is important to realize that the WERs for clean speech $x$ are supposed to be the best possible ASR results for any SE model output, regardless of input type. In addition, the perception (Perceptual Evaluation of Speech Quality, PESQ) and intelligibility (Short-Time Objective Intelligibility, STOI) scores of various SE are also reported.

Ideally, a well-designed SE model should not introduce significant changes to clean speech inputs. However, observing the CCT-AM WERs of $x$, $f^S_{org}(x)$, and $f^A_{org}(x)$, we can identify the presence of clean-conditioned artifacts $\theta_c$ introduced by the SE models. Comparing the CCT-AM WERs of $z$, $f^S_{org}(z)$, and $f^A_{org}(z)$, on the other hand, can show that the SE models reduce WERs when confronted with noisy input, yet further tweaking on SE can lead to improved outcomes. 
In this case, it is difficult to determine whether artifacts or residual noise undermines the WERs of each SE output by merely reading the statistics. As a remedy, Section~\ref{sec:dis_cs} will showcase the impact of artifacts through visualization. These observations apply to MCT-AM as well, but with a shorter performance range due to its robustness.

Furthermore, something interesting is that $f^S_{org}(x)$ exhibits worse WERs but higher PESQ than $f^A_{org}(x)$. This could demonstrate two points: 1) an SE module may not always handle a clean input appropriately, hindering its potential to adapt to new scenarios; 2) exceeding a certain level of PESQ/STOI (contribution by denoising), a higher
PESQ/STOI does not fully translate to a lower WER (deterioration by artifacts). 

\subsection{Simple enhancement}

\begin{table}[tb]
\caption{The results on simple SE. The performance superior and equal to the baseline ($f^S_{org}$) are in blue and green, respectively.}
\label{tab:simple_se}
\resizebox{\columnwidth}{!}{%
\begin{tabular}{@{}ccccccc@{}}
\toprule
 & $f^S_{pre, \alpha}(x)$ & $f^S_{pre, \alpha}(z)$ & $f^S_{pre, \beta}(x)$ & $f^S_{pre, \beta}(z)$ & $f^S_{scr, \beta}(x)$ & $f^S_{scr, \beta}(z)$ \\ \midrule
CCT-AM   & \B5.44 & \B9.63 & \B5.33 & \B9.53 & 6.36 & 14.12 \\
MCT-AM   & \B5.00 & \B7.08 & \B4.99 & \B7.03 & 6.22 & 9.03  \\ \midrule
PESQ     &   4.43 &   2.66 &   4.43 &   2.68 & 3.50 & 2.41  \\
STOI(\%) &   99.5 & \G93.8 &   99.5 & \G93.8 & 95.1 & 89.2  \\ \bottomrule
\end{tabular}%
}
\end{table}

Tab.~\ref{tab:simple_se} shows the results of employing NAaLoss on simple SE. As we can see, except for the trained-from-scratch model $f^S_{scr, \beta}$, further learning with NAaLoss makes the model outperforms the baseline $f^S_{org}$ (scores in blue) with a cost of trivial degradation in perception/intelligibility metrics. In particular, using option $beta$ with pre-trained parameters leads to the best result among various settings, approving its effectiveness of reducing artifacts in a multi-conditioned perspective. 

Furthermore, we calculate the relative WER reduction (WERR) rate to provide more insight, which is defined as:
$$
\text{WERR} = (1-\frac{\text{WER}_{NAa} - \text{WER}_{uc}}{\text{WER}_{org} - \text{WER}_{uc}})\times 100\%,
$$
where $\text{WER}_{uc}$, $\text{WER}_{org}$, and $\text{WER}_{NAa}$ are the WERs of unprocessed clean speech, the speech enhanced with the original SE model, and the speech enhanced with the NAaLoss-adopted SE. A higher WERR score signifies that applying NAaLoss can further reduce the noise and artifacts left over from the original SE model. 

When the input is $x$, using the WERs of $x$, $f^S_{org}(x)$, and $f^S_{pre, \beta}(x)$ to compute WERRs gives 33.3\% and 51.2\% for CCT-AM and MCT-AM, respectively. The WERRs of the input $z$ for CCT-AM and MCT-AM are 61.4\% and 53.9\%, respectively, using the WERs of $x$, $f^S_{org}(z)$, and $f^S_{pre, \beta}(z)$.
We have found that the presented NAaLoss helps the simple SE achieve better ASR results, and even helps more with noisy speech $z$ than with clean speech $x$ (33.3\% to 61.4\% and 51.2\% to 53.9\% in WERR). This attributes to individual modeling of speech, artifacts, and noise in NAaLoss. Furthermore, since MCT-AM outperforms CCT-AM in ASR, using NAaLoss to reduce artifacts in enhanced clean speech $f^S_{pre, \beta}(x)$ can achieve higher WERR in MCT-AM than in CCT-AM (51.2\% vs. 33.3\% in WERR). 

\subsection{Advanced enhancement}

\begin{table}[tb]
\caption{The results on advanced SE. While the color settings are the same in Tab.~\ref{tab:simple_se}, those WERs better than the performance of clean speech $x$ are in red.}
\label{tab:advanced_se}
\resizebox{\columnwidth}{!}{%
\begin{tabular}{@{}ccccccc@{}}
\toprule
 & $f^A_{pre, \alpha}(x)$ & $f^A_{pre, \alpha}(z)$ & $f^A_{pre, \beta}(x)$ & $f^A_{pre, \beta}(z)$ & $f^A_{scr, \beta}(x)$ & $f^A_{scr, \beta}(z)$ \\ \midrule
CCT-AM   & \B5.16 & \B7.06 & \B5.17 & \B6.83 & \R5.02 &   7.45 \\
MCT-AM   & \B4.89 & \B6.39 & \B4.88 & \B6.41 & \R4.81 & \B6.46 \\ \midrule
PESQ     &   4.18 & \B3.13 & \G4.22 & \G3.11 &   4.21 &   2.97 \\
STOI(\%) &   98.6 &   94.6 & \G98.8 &   94.8 & \B99.0 &   94.6 \\ \bottomrule
\end{tabular}%
}
\end{table}

The results of employing NAaLoss on advanced SE can be seen in Tab.~\ref{tab:advanced_se}. Excluding the case of CCT-AM for $f^A_{scr, \beta}(z)$, the advanced SE consistently achieves lower WERs (scores in blue) if it is further adopted by NAaLoss.  Again, using option $\beta$ with pre-trained parameters guides to the best result on average (blue in WER, green in PESQ/STOI), confirming the stability of this arrangement. Additionally, we spot that some enhanced versions of speech get higher perception/intelligibility scores than the baseline, e.g., PESQ of $f^A_{pre, \alpha}(z)$ and STOI of $f^A_{scr, \beta}(x)$, hinting at the synergistic capacity of integrating MANNER and NAaLoss in SE. Interestingly, $f^A_{scr, \beta}(x)$ obtains lower WERs than clean speech $x$ (5.02\% vs. 5.04\% under CCT-AM and 4.81\% vs. 4.86\% under MCT-AM), showing that this SE setup can further benefit clean speech in ASR.
\comment{Interestingly, although $f^A_{scr, \beta}(x)$ obtains lower WERs than $x$ (scores in red), it has strayed from its principal task, denoising, which explains its subordinate result under CCT-AM.}

Similarly, we use the WERR metric to evaluate the effect of NAaLoss on advanced SE. 
The WERRs for CCT-AM and MCT-AM are 45.8\% and 60.0\% in input $x$, and 23.2\% and 11.9\% in input $z$.
Again, for the clean speech input $x$, using NAsLoss can decrease the artifacts caused by the advanced SE, resulting in a more significant reduction in WERs for MCT-AM than for CCT-AM (60\% vs. 50\%), a tendency also found in the simple SE case. However, NAaLoss contributes less to noisy speech input $z$ than to clean speech $x$ in WERR (23.2\% vs. 50.0\% for CCT-AM and 13.1\% vs. 60.0\% for MCT-AM), particularly for MCT-AM. The underlying explanation could be that MCT-AM can deal with the artifact and residual noise left by advanced SE to some extent, and the effect of utilizing NAaLoss is less significant.

\subsection{Comparison}

\begin{table}[tb]
\caption{The comparison between OA and NAaLoss.}
\label{tab:oa_comp}
\resizebox{\columnwidth}{!}{%
\begin{tabular}{ccccc}
\toprule
       & $f^S_{org}(z)+0.5z$ & $f^S_{pre, \beta}(z)$ & $f^A_{org}(z)+0.5z$ & $f^A_{pre, \beta}(z)$ \\ \midrule
CCT-AM & 15.01  & \B9.53 & 8.84 & \B6.83 \\
MCT-AM & \B7.09 & \B7.03 & 6.92 & \B6.41 \\ \bottomrule
\end{tabular}%
}
\end{table}

Here, we evaluate the observation-adding (OA) method \cite{artif} and compare it with our presented NAaLoss-wise framework, which WER results are listed in Tab.~\ref{tab:oa_comp}. OA is utilized in multiple research fields \cite{oa1,oa2}, with the intuition to lessen nonlinear audio distortion, such as artifacts. Simply adding a portion of the noisy speech $z$ to the enhanced speech $f(z)$ defines the OA method. What we have observed in Tab.~\ref{tab:oa_comp} is fourfold:
1) NAaLoss outperforms OA in each circumstance;
2) for CCT-AM, which is sensitive to noise, adding back noise as in OA is destructive (causing WER from 10.21\% to 15.01\% for simple SE and from 7.37\% to 8.84\% for advanced SE);
3) the effectiveness of OA is counterproductive on advanced SE, compared to baselines (8.84\% v.s. 7.37\% for CCT-AM, and 6.92\% v.s. 6.62\% for MCT-AM);
4) in the instance of $f^S_{org}(z)+0.5z$ with MCT-AM, OA performs closest to NAaLoss (7.09\% vs. 7.03\% in WER), which coheres with the experiments in \cite{artif}. Although OA is beneficial to a noise-robust ASR, it may not serve as a SE front-end since it could deteriorate the quality and intelligibility of enhanced speech. In comparison, NAaLoss fine-tunes the SE model parameters with the objective of minimizing artifacts while maintaining enhancement quality in various scenarios, accentuating its comprehensiveness.

\section{Discussion}
\label{sec:dis_cs}

\begin{figure}[h]
\begin{minipage}[b]{1.0\linewidth}
  \centering
  \centerline{\epsfig{figure=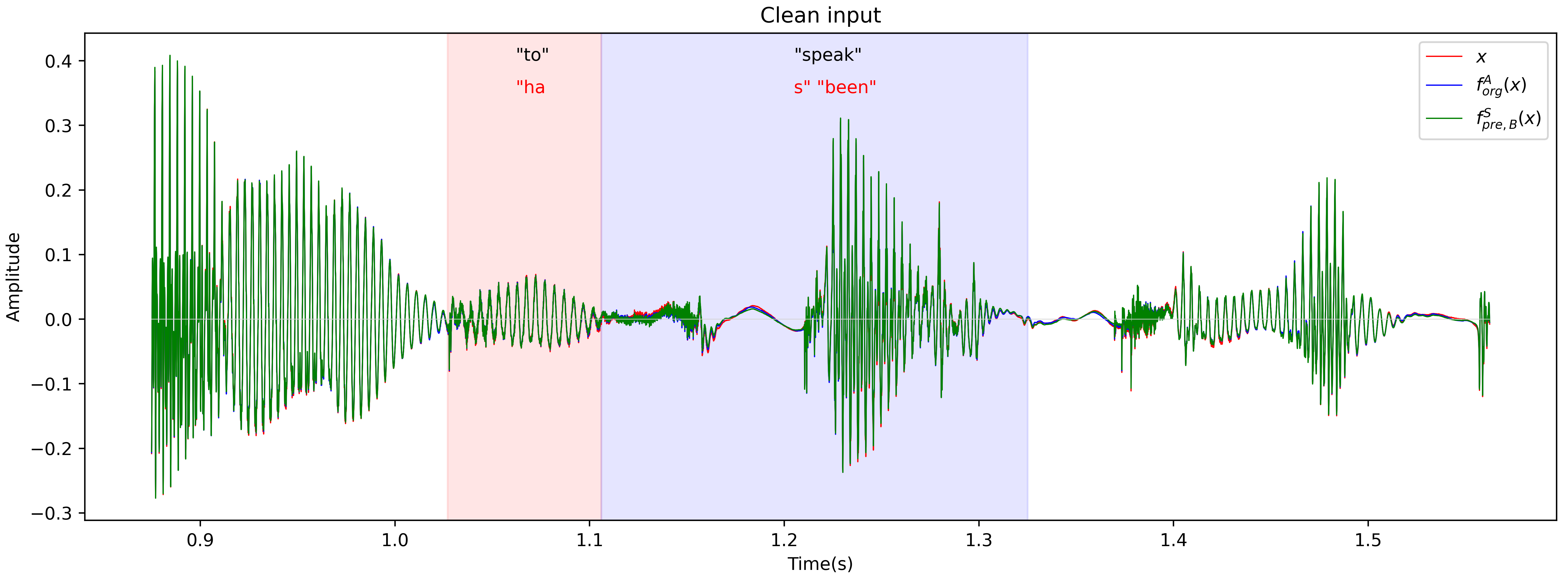, width=8cm}}
   \centerline{\small{(a) Overall waveform of clean input. Clean speech $x$ are in red.}}\medskip
\end{minipage}
\begin{minipage}[b]{1.0\linewidth}
  \centering
  \centerline{\epsfig{figure=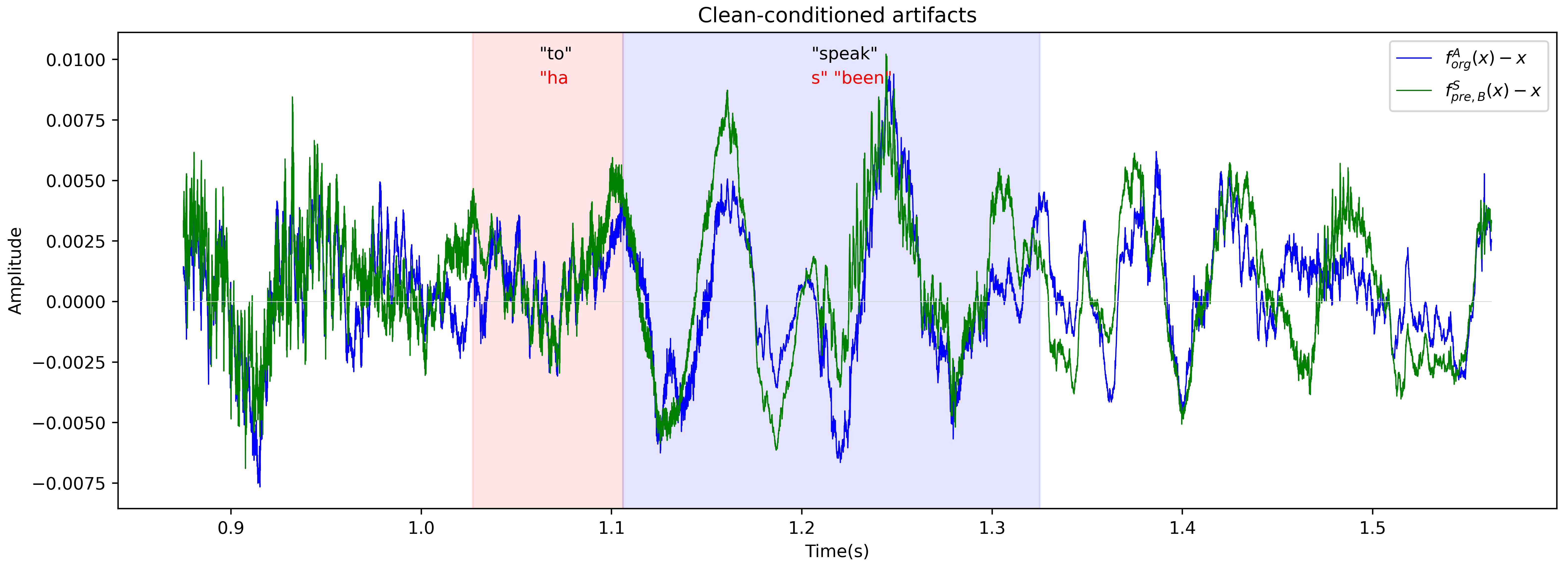, width=8cm}}
  \centerline{\small{(b) Clean-conditioned artifacts $\theta_c$.}}\medskip
\end{minipage}
\begin{minipage}[b]{1.0\linewidth}
  \centering
  \centerline{\epsfig{figure=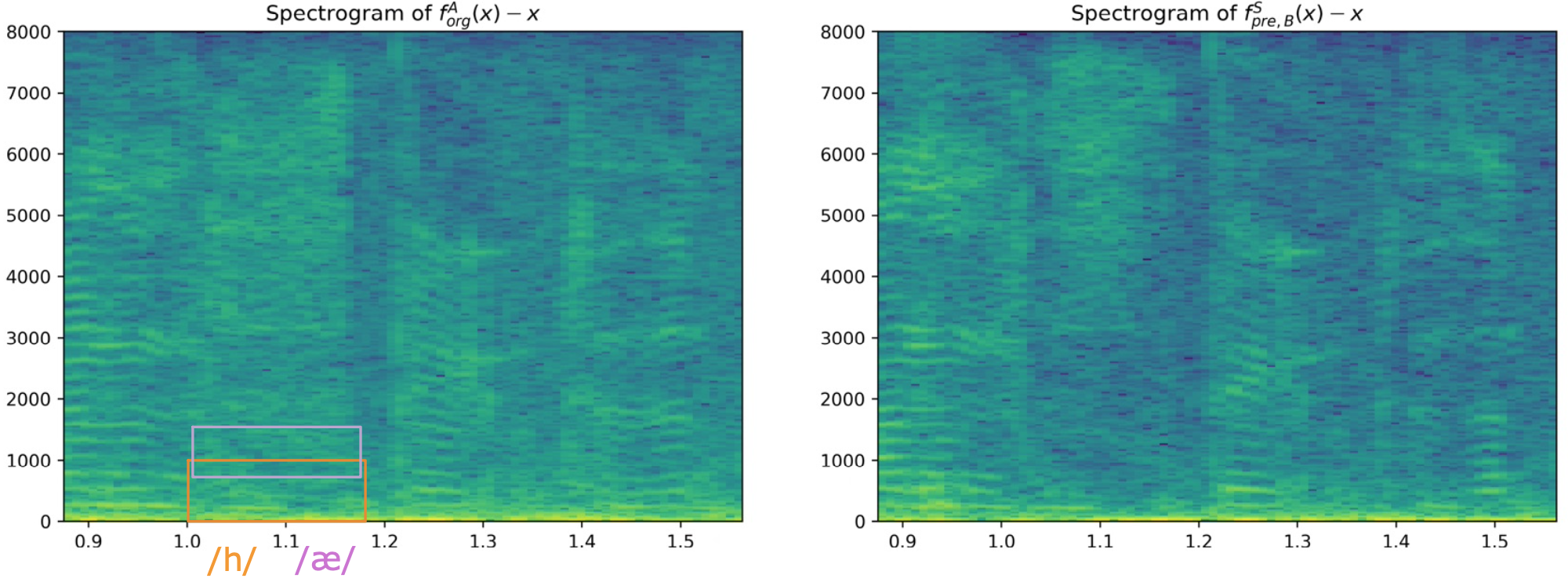, width=8cm}}
  \centerline{\small{(c) Spectrogram of $\theta_c$ in $f^A_{org}(x)$ (left), and in $f^A_{pre, \beta}(x)$ (right).}}\medskip
\end{minipage}
\caption{Case studies on Hypothesis 1. 
This is an example of the ground truth "to speak" recognized correctly in $f^A_{pre, \beta}(x)$ but falsely as "has been" in $f^A_{org}(x)$. Blue and green lines denote production related to $f^A_{org}(x)$ and $f^A_{pre, \beta}(x)$, respectively. For sub-figure (a) and (b), we underline the timeline of the respective phoneme, and the false recognitions are in red. The bounding boxes in (c) indicate the frequency range of the specified phoneme.}
\label{fig:case_clean}
\end{figure} 

\begin{figure}[h!]
\begin{minipage}[b]{1.0\linewidth}
  \centering
  \centerline{\epsfig{figure=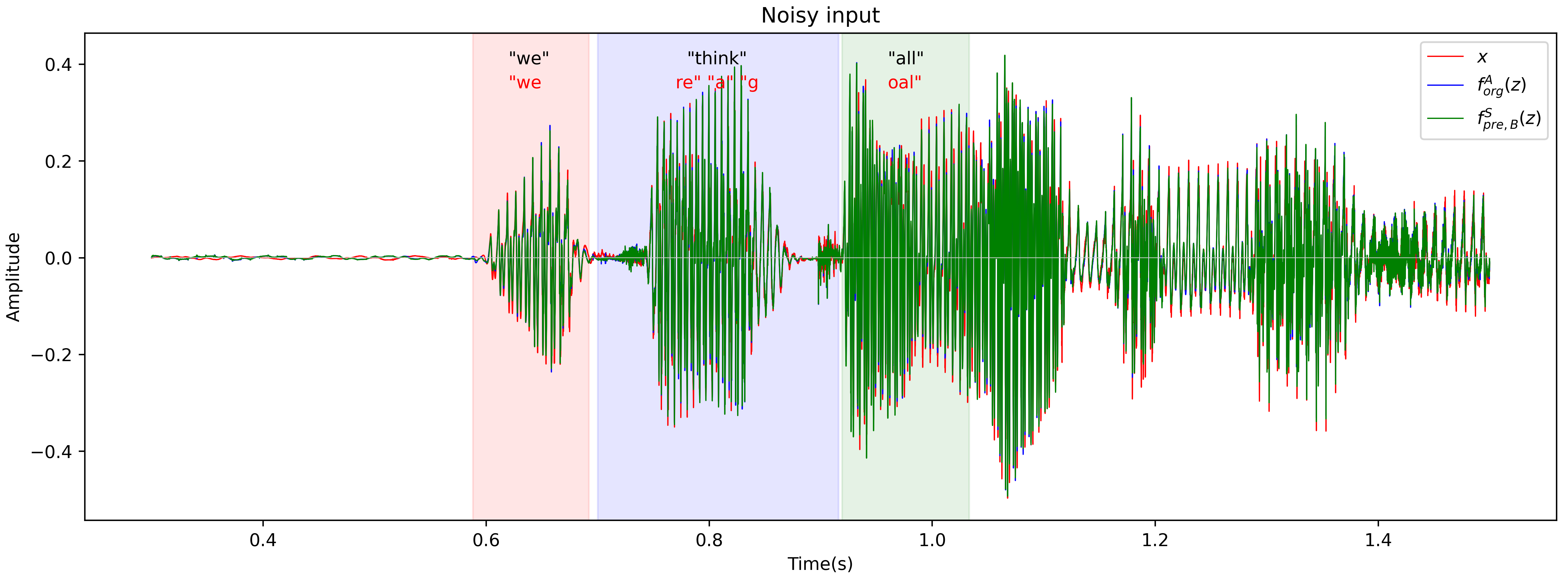, width=8cm}}
   \centerline{\small{(a) Overall waveform of noisy input.}}\medskip
\end{minipage}
\begin{minipage}[b]{1.0\linewidth}
  \centering
  \centerline{\epsfig{figure=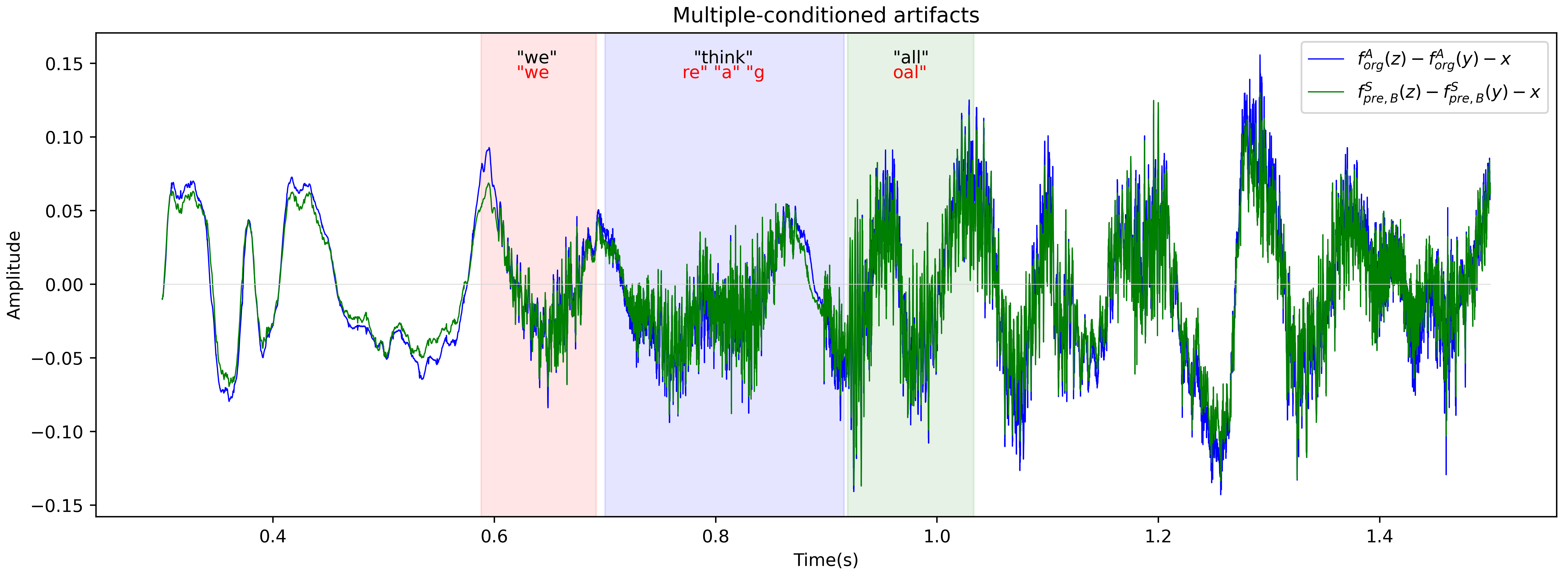, width=8cm}}
  \centerline{\small{(b) Multi-conditioned artifacts $\theta_m$.}}\medskip
\end{minipage}
\begin{minipage}[b]{1.0\linewidth}
  \centering
  \centerline{\epsfig{figure=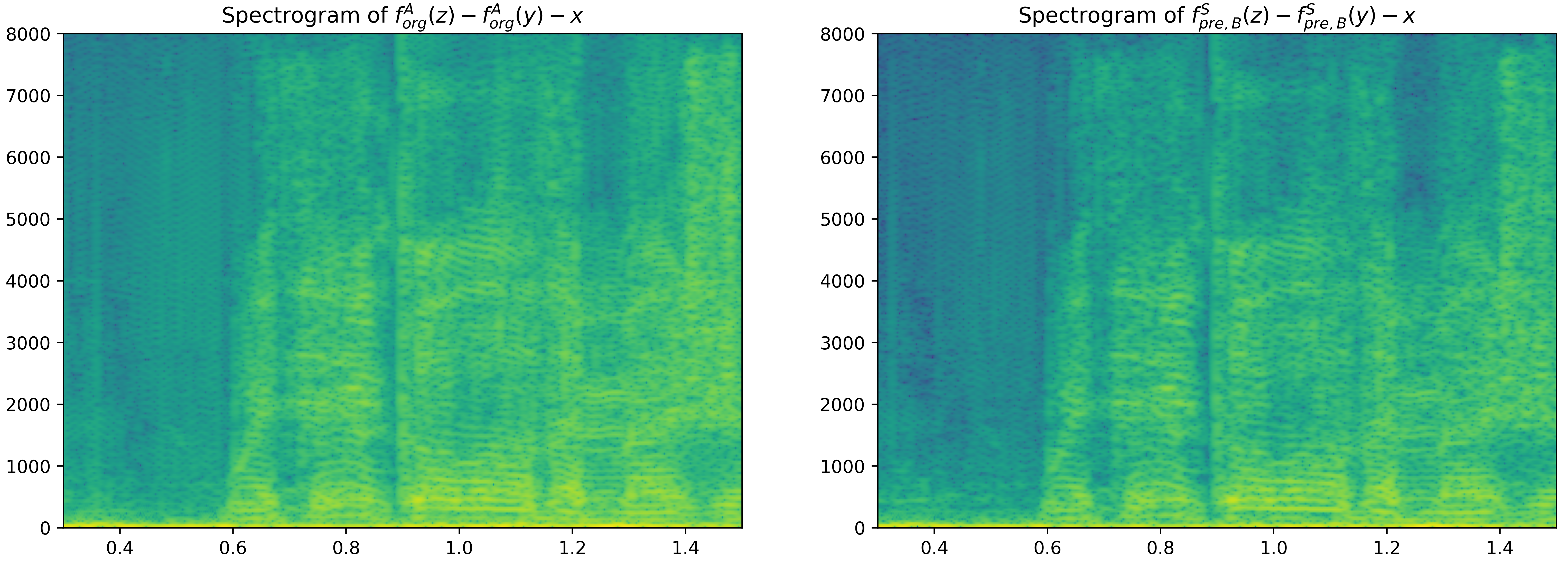, width=8cm}}
  \centerline{\small{(c) Spectrogram of $\theta_m$ in $f^A_{org}(z)$ (left), and in $f^A_{pre, \beta}(z)$ (right).}}\medskip
\end{minipage}
\begin{minipage}[b]{1.0\linewidth}
  \centering
  \centerline{\epsfig{figure=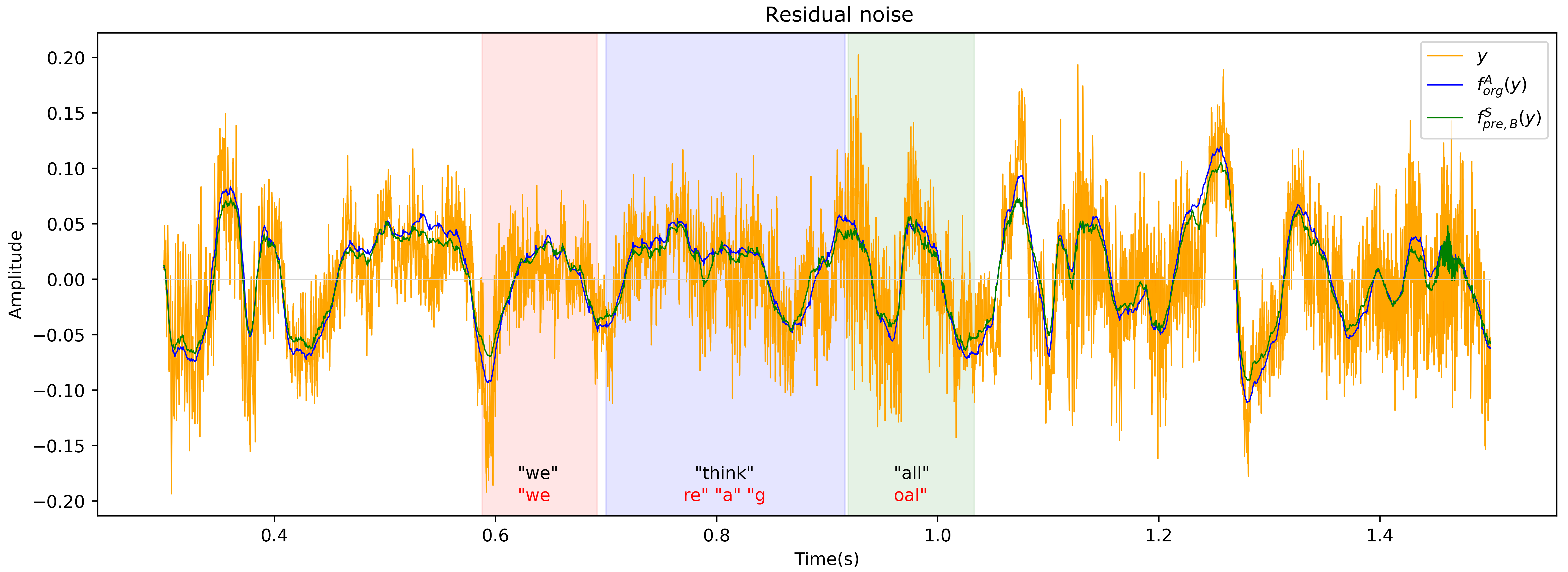, width=8cm}}
  \centerline{\small{(d) Residual noise $\tilde{y}$. The waveform of pure noise $y$ is in yellow.}}\medskip
\end{minipage}
\caption{Case studies on Hypothesis 2 and 3. This is an example of the ground truth "we think all" recognized correctly in $f^A_{pre, \beta}(z)$ but falsely as "were a goal" in $f^A_{org}(z)$. The color settings are identical to that of Fig.~\ref{fig:case_clean}.}
\label{fig:case_multi}
\end{figure} 

This section seeks to further analyze the characteristics of the presented NAaLoss in light of the evaluation results offered in the previous section. Regarding the outcomes of simple and advanced SE, it appears necessary to compromise on some perception/intelligibility scores in order to achieve a higher ASR result (a lower WER). However, since NAaLoss considers multiple aspects within three constituent losses described in Section \ref{sec:formulation}, the resulting SE model is supposed to provide a better trade-off between SE and ASR performance. Moreover, it is widely accepted that perceptually degraded speech has little effect on human recognition, albeit the human auditory system is susceptible to noise. 
Furthermore, we find that the perception/intelligibility scores of $f^A_{pre, \beta}(x)$ are lower than that of $f^S_{pre, \beta}(x)$, implying that a more complicated SE model tends to be less capable of handling clean speech input.
To mitigate this problem, we suggest tuning the weight for the loss of de-artifact ($L_{\text{deatf}}$) in NAaLoss, especially in the clean-conditioned scenario, whereas we leave this to future works.

Regarding the transcription results, we further analyze some instances of wrong-recognized words in $f^A_{org}$ and explicitly reveal the impact of artifacts on ASR. First, in the case of input $x$, $f^A_{org}$ tends to wipe out or change the consonants, e.g., "hores" misrecognized as "ores" or "if" misrecognized as "is." Second, in the case of input $z$, multiple samples have reported misaligned and thus misrecognized words because of the artifacts that alter the timing of speech. For example, "the same" is misaligned by timing artifacts, leading to misrecognition as "*** plain," where "***" denotes an unknown symbol. 

Since enumerating all the WERs is impossible, we visualize two samples, one analyzing hypothesis 1 and another analyzing hypotheses 2 and 3, as shown in Fig.~\ref{fig:case_clean} and Fig.~\ref{fig:case_multi}, respectively. As Fig.~\ref{fig:case_clean} (a) displays, the waveforms of $x$, $f^A_{org}(x)$, and $f^A_{pre, \beta}(x)$ are too close to tell the difference; however, plotting the clean-conditioned artifacts, as in Fig.~\ref{fig:case_clean} (b), it is evident that both signals have distinct characteristics. We then draw the respective spectrogram in Fig.~\ref{fig:case_clean} (c) to better identify their frequency components and annotate the misrecognized part. Because some artifacts (1.1$\sim$1.3 s) distribute on the primary frequency ranges of consonant /h/ (0$\sim$1000 Hz) and vowel /æ/ (800$\sim$1500 Hz) \cite{consonant1, consonant2}, they interfere with the ASR to choose "has" rather than "to." This also affects the language model in ASR to select the next word, "been," which is reasonable but wrong. Additionally, we can observe that artifacts distribute widely and densely in Fig.~\ref{fig:case_clean} (c) and Fig.~\ref{fig:case_multi} (c), which may give rise to the false transcription. Last, we plot the residual noise in Fig.~\ref{fig:case_multi} (d) and show that NAaLoss is also better in noise reduction as $f^A_{pre, \beta}(y)$ contains smaller residual noise than $f^A_{org}(y)$.

\section{Conclusion}
\label{sec:conclu}
This study proposes a novel objective function NAaLoss adapting SE models toward ideal performance. We experimentally reveal the effectiveness of NAaLoss in 1) eliminating artifacts, 2) enhancing noisy speech, 3) preserving perceptual quality and intelligibility, and 4) generalizing to subsequent ASR of any form. The waveforms and their corresponding spectral visualizations to show the effectiveness of NAaLoss can be found in supplementary files. In the future, we plan on creating an automatic mechanism to determine the weights of the three constituent losses in NAaLoss. We also intend to encourage more instantiations in future research that studies single-channel SE front-ends considering the specific requirements and constraints of the ASR task. 


\vspace{12pt}

\end{document}